\documentclass[12pt]{iopart}
\usepackage{iopams,graphicx}  
\begin{document}

\title[letter]{Transport Coefficients of the Anderson Model}

\author{T.\ A.\ Costi$^1$\footnote{Present address:
Peter Gr\"unberg Institute, Research Centre J\"ulich, 52425 J\"ulich, Germany}  
and A.\ C.\ Hewson$^1$}

\address{$^1$ Department of Mathematics, 
Imperial College, London SW7 2BZ, UK}
\begin{abstract}
The transport coefficients of the Anderson model require knowledge of both the
temperature and frequency dependence of the single--particle spectral
densities and consequently have proven difficult quantities to
calculate. Here we show how these quantities can be calculated via an 
extension of Wilson's numerical renormalization group method. Accurate
results are obtained in all parameter regimes and for the full range
of temperatures of interest ranging from the high temperature
perturbative regime $T>>T_{K}$, through the cross--over region $T\approx
T_{K}$, and into the low temperature strong coupling regime $T<<T_{K}$.
The Fermi liquid relations for the $T^2$ coefficient of the
resistivity and the linear coefficient of the thermopower are
satisfied to a high degree of accuracy. The techniques used here provide a
new highly accurate approach to strongly correlated electrons in high
dimensions.

\end{abstract}

\maketitle

\section{Introduction}
In this Letter we present accurate new results for the transport
coefficients of the Anderson model obtained by using the numerical renormalization
group method \cite{wilson75}. The Anderson model has been used
extensively to interpret the properties of magnetic impurities in
simple metals, certain aspects of heavy fermion behaviour
\cite{hewson93} and the low--temperature transport
through a quantum dot \cite{meir93}. Recently this model has also become important for an understanding
of the properties of correlated electrons on a lattice in high
dimensions. The electron self--energy becomes local in infinite
dimensions \cite{mh89,metzner89} and the problem reduces to that of an
impurity \cite{mielsch_all}. In particular the solution of the $d=\infty$ Hubbard model has been
shown to reduce to the solution of an Anderson impurity model with a hybridization
function, $\Delta(\omega)$, determined by a self--consistency
requirement \cite{george92,janis91,janis92}. The self--consistency
condition is straightforward to handle 
and the difficult part of the problem has
remained the calculation of the dynamics of the underlying impurity
problem, and more specifically of the impurity single--particle
spectral densities. 
This is just the problem which one encounters in attempting to
calculate the transport 
coefficients of the Anderson model which depend sensitively on the
temperature and frequency dependence of the single--particle spectral
densities $\rho_{d}(\omega,T)$. An accurate solution of this problem can
therefore provide the solution of the $d=\infty$ Hubbard model. 
Our main interest, then, is the accurate solution of the Anderson
model and in particular the calculation of its single--particle
spectral densities.
The Bethe Ansatz has been successful in providing exact results on the
thermodynamic properties of several models of magnetic impurities
including the N--fold degenerate Anderson model
\cite{andrei83,tsvelick83}, but the method cannot be used to extract dynamic
properties. The latter have been obtained by applying approximate
methods. For  $N=2$ perturbation theory in the local Coulomb repulsion, U, has
been used \cite{horvatic87} whilst for larger N (typically $N\ge 6$) the
non--crossing approximation (NCA) has been applied to calculate the
thermodynamic,  
transport and dynamic properties \cite{bickers87,qin91}. The former eventually becomes
unreliable for large enough values of U (typically for $U/\pi\Delta
\ge 2.5$ where $\Delta$ is the resonance level width) and
the latter fails to satisfy some exact Fermi liquid relations at low
energies \cite{mh84}. Recently, Quantum Monte Carlo methods
in combination with the maximum entropy principle have also been used to
extract dynamic properties \cite{silver90,jarrell91a,jarrell91b}. These have so far
been restricted to the symmetric model and to $U/\pi\Delta \le 3$.
They become more inaccurate and time consuming for larger values of U and
lower temperatures.
\subsection{Model and method}
A comprehensive and highly accurate approach to the calculation of 
dynamic properties of magnetic impurity models has recently been
developed, based on an extension of Wilson's numerical renormalization
group approach, which overcomes many of the limitations and
difficulties of the other approximate schemes. This has led to
reliable results valid in all regimes and at both zero 
\cite{costi90,costi91,costi92a,frota86,sakai89,sakai90} and finite
\cite{costi92b} temperature, and which satisfy all the sum rules and 
Fermi liquid relations of the Anderson model \cite{costi91}.
With such results for the dynamic properties it 
becomes possible to calculate the transport coefficients of the
Anderson model over the full temperature range of interest. We report
here results for the resistivity, and thermopower based on these
numerical renormalization 
group calculations. Results for the Hall coefficient and thermal
conductivity have also been calculated and will be presented in a forthcoming
publication \cite{costi93b}.

Our starting point is the non--degenerate ($N=2$) Anderson model:
\begin{eqnarray}
H & = & \sum_{\sigma}\epsilon_{d}c_{d\sigma}^{\dagger}c_{d\sigma} +
Un_{d\uparrow}n_{d\downarrow} +
\sum_{k\sigma}(V_{k\sigma}c_{k\sigma}^{\dagger}c_{d\sigma} + 
H.c.) + \sum_{k\sigma}\epsilon_{k}c_{k\sigma}^{\dagger}c_{k\sigma}
\end{eqnarray}
This can be re--cast in the form of a linear chain model which may
then be iteratively diagonalized. The Lanczos algorithm is applied to
$H_{c}=\sum_{k\sigma}\epsilon_{k}c_{k\sigma}^{\dagger}c_{k\sigma}$ 
with starting vector $c_{0\sigma}^{\dagger}|vac>={1\over
V}\sum_{k}V_{k}c_{k\sigma}^{\dagger}|vac>$ where
$V=\sqrt{\sum_{k}|V_{k}|^{2}}$ so that the conduction electron part
of $H$ is reduced to a semi--infinite linear chain with site $0$ coupled
to the impurity via a hybridization term of strength V \cite{wilson75,costi90}:
\begin{eqnarray}
H & = & \sum_{\sigma}\epsilon_{d}c_{d\sigma}^{\dagger}c_{d\sigma} +
Un_{d\uparrow}n_{d\downarrow} +
V\sum_{\sigma}(c_{0\sigma}^{\dagger}c_{d\sigma} + H.c.)\nonumber\\
 & + &
\sum_{n=0,\sigma}^{\infty}\epsilon_{n}c_{n\sigma}^{\dagger}c_{n\sigma} +
\sum_{n=0,\sigma}^{\infty}\lambda_{n}(c_{n+1\sigma}^{\dagger}c_{n\sigma}+c_{n\sigma}^{\dagger}c_{n+1\sigma})
\end{eqnarray}
The parameters $\{ \epsilon_n, \lambda_n\}, n=0,1,\ldots$ reflect the
form of the hybridization function $\Delta(\omega)=\pi
\sum_{k}|V_{k}|^2 \delta(\omega - \epsilon_{k})$. In the present case
of the magnetic 
impurity problem we are primarily interested in low energies where
$\Delta(\omega)$ may be taken to be a slowly
varying function of $\omega$ which can be approximated by a constant $\Delta$.
In this case, and by replacing 
the continuum of conduction electron states in $(-D,D)$ by a discrete
set $\pm D\Lambda^{-n}, n=0,1,\ldots, \Lambda > 1$, a discrete
approximation to $H$ can be obtained with parameters
$\lambda_{n} \sim D\Lambda^{-n/2}, n >> 1$ \cite{wilson75} (the
$\epsilon_{n}$ are zero for a half--filled symmetric band). 
For the impurity model corresponding to the $d=\infty$
Hubbard model, the frequency dependence of $\Delta(\omega)$ is
important and leads to a different set of parameters $\lambda_{n}$.
The model has to be solved repeatedly in this case to obtain a
self--consistent $\Delta(\omega)$ using the self--consistency
requirement in \cite{george92,janis91,janis92}.

 The discretized Anderson model with $\lambda_{n} \sim D\Lambda^{-n/2},
n >> 1$, discussed above, can be iteratively diagonalized to obtain the
many--body eigenvalues $E_{p}^{n}$ and 
eigenstates $|p>_{n}$ on successively lower energy scales $\omega_{n}
\sim D\Lambda^{-n/2}, n=0,1,\ldots$ by following the procedure in
\cite{wilson75}. The 
procedure starts with diagonalizing the impurity part
$H_{0}=\sum_{\sigma}\epsilon_{d}c_{d\sigma}^{\dagger}c_{d\sigma} +
Un_{d\uparrow}n_{d\downarrow}$ and then adding the
coupling to the local conduction electron state
$V\sum_{\sigma}(c_{0\sigma}^{\dagger}c_{d\sigma} + H.c.)$. After this
one adds successive energy shells
$\sum_{\sigma}\lambda_{n}(c_{n+1\sigma}^{\dagger}c_{n\sigma}+c_{n\sigma}^{\dagger}c_{n+1\sigma}),
n=0,1,\ldots $ and diagonalizes the resulting Hamiltonian. 
Since the number of states grows like $4^{n}$ it is 
in practice only possible to retain the lowest 500 or so states at
each stage. This restricts the reliable range of excitations $\omega$
to be such that $\omega_{n} \leq \omega \leq K\omega_{n}$, where $K\approx 10$
for $\Lambda \sim 3$. The lower excitations are calculated more
reliably in successive iterations, whilst information on the higher
excitations is contained in previous iterations. 
The eigenvalues are used to calculate the partition function at a
decreasing sequence of temperatures $T_{n} = \omega_{n}/k_{B} \sim D\Lambda^{-n/2},
n=0,1,\ldots$ and from this one extracts the thermodynamic properties \cite{wilson75,krishnamurthy80}.
\section{\bf Results}
\subsubsection{\bf Spectral functions} 
From the eigenvectors $|p>_{n}$ we recursively
evaluate the single--particle matrix elements
$M_{pp'}^{n}=<p|c_{d\sigma}|p'>$ which are required for the n'th shell Green's
function $G_{d\sigma}^{n}(\omega,T)$
and spectral density $\rho_{d}^{n}(\omega,T)$,
\begin{equation}
G_{d\sigma}^{n}(\omega,T) = <<c_{d\sigma};c_{d\sigma}^{\dagger}>> 
={1\over{Z_{n}(\beta)}}\sum_{pp'}{{{|M_{pp'}^n|}^2{(\thinspace e^{-\beta
E_{p}^n}+e^{-\beta E_{p'}^{n}}\thinspace)}}\over{\omega - E_{p'}^n + E_{p}^n}}
\end{equation}
\begin{equation}
\rho_{d}^{n}(\omega,T)  = {1\over{Z_{n}(\beta)}}\sum_{pp'}{{{|M_{pp'}^n|}^2 
(\thinspace e^{-\beta E_{p}^n}+e^{-\beta
E_{p'}^{n}}\thinspace)\delta(\omega - E_{p'}^n + E_{p}^n}}) 
\end{equation}
Here $Z_{n}(\beta)$ is the partition function for the $n'th$
cluster\footnote{The term {\em cluster} is suggested by the notation,
although the calculations described here are in k--space
\cite{wilson75}} or energy shell.
In evaluating $\rho_{d}^{n}(\omega,T)$ the delta functions
are replaced by 
Gaussians having widths $\alpha_{n}$ of order $\omega_{n}$ appropriate to the
cluster size. At $T=0$ the spectral density is evaluated at excitations $\omega
\approx \omega_{n}$ using the $n'th$ cluster. At finite temperature
$k_{B}T>0$, and when $\omega\approx\omega_{n}$ becomes of order
$k_{B}T$ excited states not contained in the $n'th$ cluster become important. In this case we
use a smaller cluster (containing information about the higher
excited states) to evaluate the spectral densities at energies
$\omega$ such that $\omega \leq k_{B}T$. Results for the spectral
densities $\rho_{d}(\omega,T)$ are shown in Figure~\ref{figure1}
and Figure~\ref{figure2} 
for various temperatures.

The Friedel sum rule
$\rho_{d}(\omega=\epsilon_{F},T=0)={1 \over {\pi\Delta}}\sin^{2}(\pi
n_{d}/2)$ and Shiba relation for the dynamic susceptibility \cite{shiba75} are
satisfied to within a few per cent in all parameter regimes
\cite{costi90,costi91,costi93c}. 

\subsubsection{\bf Transport coefficients}  
The calculation of transport properties is straightforward once
accurate results for the temperature and frequency dependence of the
spectral densities are obtained. The resistivity $\rho(T)$ and
thermopower $S(T)$ are given in terms of the integrals
\begin{equation}
L_{ml}=-C\int_{-\infty}^{+\infty}{\partial f(\omega) 
\over \partial \omega} \tau^{l}(\omega){\omega}^{m}d\omega
\end{equation}
by
\begin{equation}
\rho(T) = {1 \over {e^{2} L_{01}}}
\end{equation}
\begin{equation}
S(T) = -{1 \over {e T}}{L_{11} \over L_{01}}
\end{equation}
where the relaxation time $\tau(\omega,T)$ is given in terms of the
local d--electron Green's function by $\tau(\omega,T)^{-1}=2\pi
|V|^{2}\rho_{d}(\omega,T)$, and $e$ is the electronic
charge. There are two exact Fermi liquid relations for the transport
coefficients. The first is for the coefficient of the $T^2$ term in the
low temperature resistivity \cite{nozieres74,yamada75}. For $T<<T_{K}$
and in the Kondo regime we have
\begin{equation}
\rho(T) = \rho(0)\left\{ 1 - c \left( {T \over T_{K} } \right)^{2}\right\}
\end{equation}
where $c = {{\pi}^4 \over 16} = 6.088$, and $T_{K}$
is defined by
\begin{equation}
k_{B}T_{K} = U{\left(\Delta\over 2U \right )^{1/2}}e^{\pi
\epsilon_{d}(\epsilon_{d}+U)/{2\Delta U}}
\end{equation}
Results for the resistivity are shown in Figure~\ref{figure3} for $U/\pi\Delta=4$ and
for several values of the local level position ranging from the Kondo
regime ($\epsilon_{d}/\Delta=-{U \over 2\Delta}, -4, -3, -2$) to the mixed
valent regime ($\epsilon_{d}/\Delta=-1,0$) and the empty orbital
regime ($\epsilon_{d}/\Delta=+1$). 
For high temperatures $T>>T_{K}$,
as described elsewhere \cite{costi92b}, the resistivity is well
described by the Hamann result \cite{hamann67}.
In the Fermi liquid regime $T<<T_{K}$ the
inset in Figure~\ref{figure3} shows the expected $T^{2}$ behaviour. In the Kondo
regime the values of the $T^{2}$ coefficient extracted from a least
squares fit in the region $0\le T \le 0.1T_{K}$ are $5.7, 5.8, 6.4$ and
$6.6$ in going from the symmetric case to $\epsilon_{d}=-2\Delta$.
These values agree to within 8\% of the exact result $c=6.088$.
Quantum Monte Carlo results for the symmetric case for $U/\pi\Delta
\le 3$ give the
$T^{2}$ coefficient to within 19\% of the exact result \cite{jarrell91b}. The current
approach which is not restricted to very low or very high temperatures
also gives accurate results for the resistivity in the cross--over
region $T \sim T_{K}$ \cite{costi92b}. The $T=0$ resistivity $\rho(0)$
satisfies $\rho(0) \sim \sin^{2}(\pi n_{d}/2)$ in accordance with the
Friedel sum rule for the single--particle spectral density \cite{costi93b}.
\begin{figure}[t]
\includegraphics[width=\linewidth,clip]{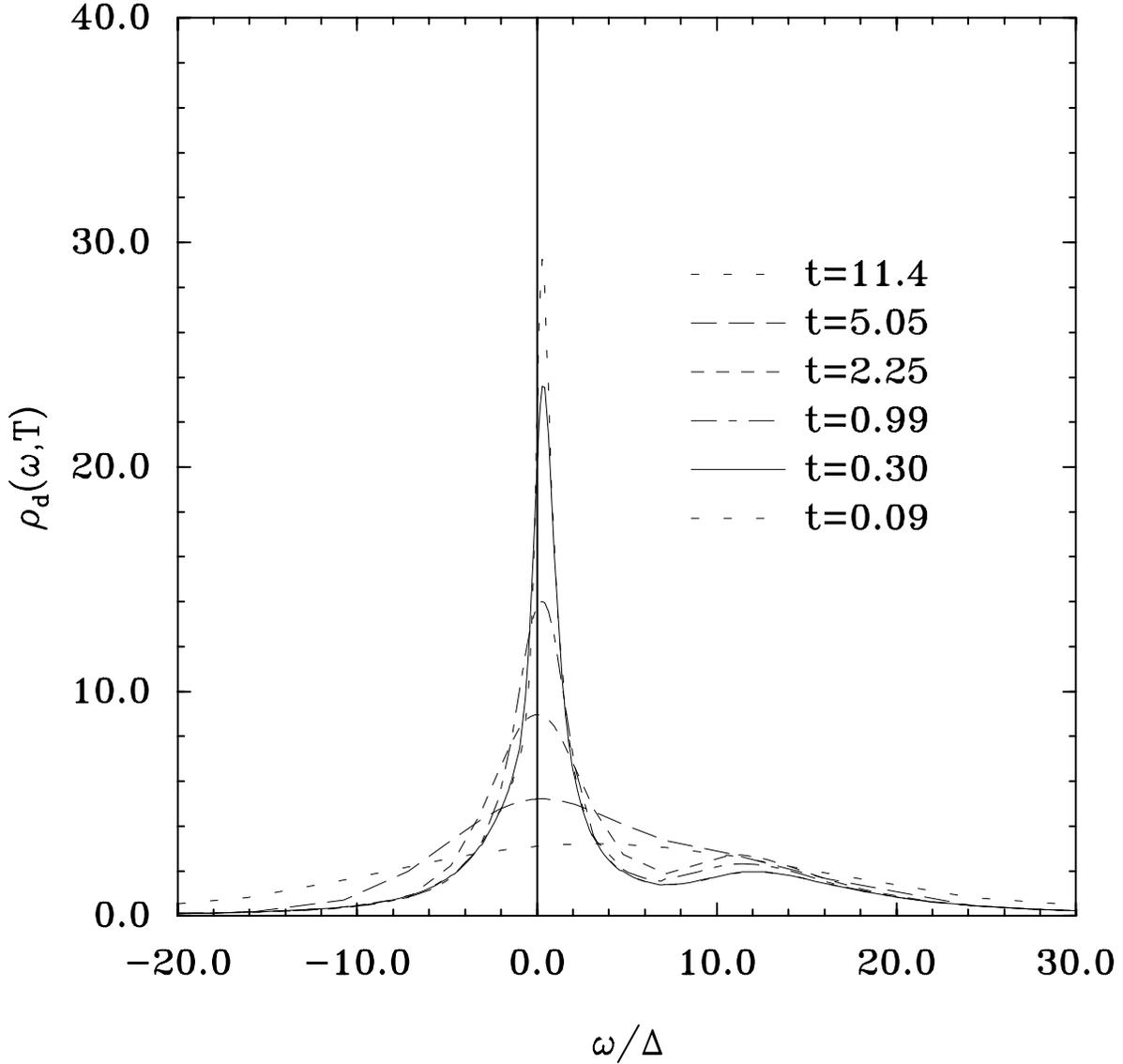}
\caption{The spectral density $\rho_{d}(\omega,T)$ in the mixed valent regime
$\epsilon_{d}/\Delta = -1$ for several values of the reduced temperature
$t=T/T_{K}$ and for $U/\pi\Delta=4$. Note that $\Delta$ is a more
appropriate energy scale in this regime than $T_{K}$. For the present
parameters $T_{K} = 0.59\Delta$}
\label{figure1}
\end{figure}
\begin{figure}[t]
\includegraphics[width=\linewidth,clip]{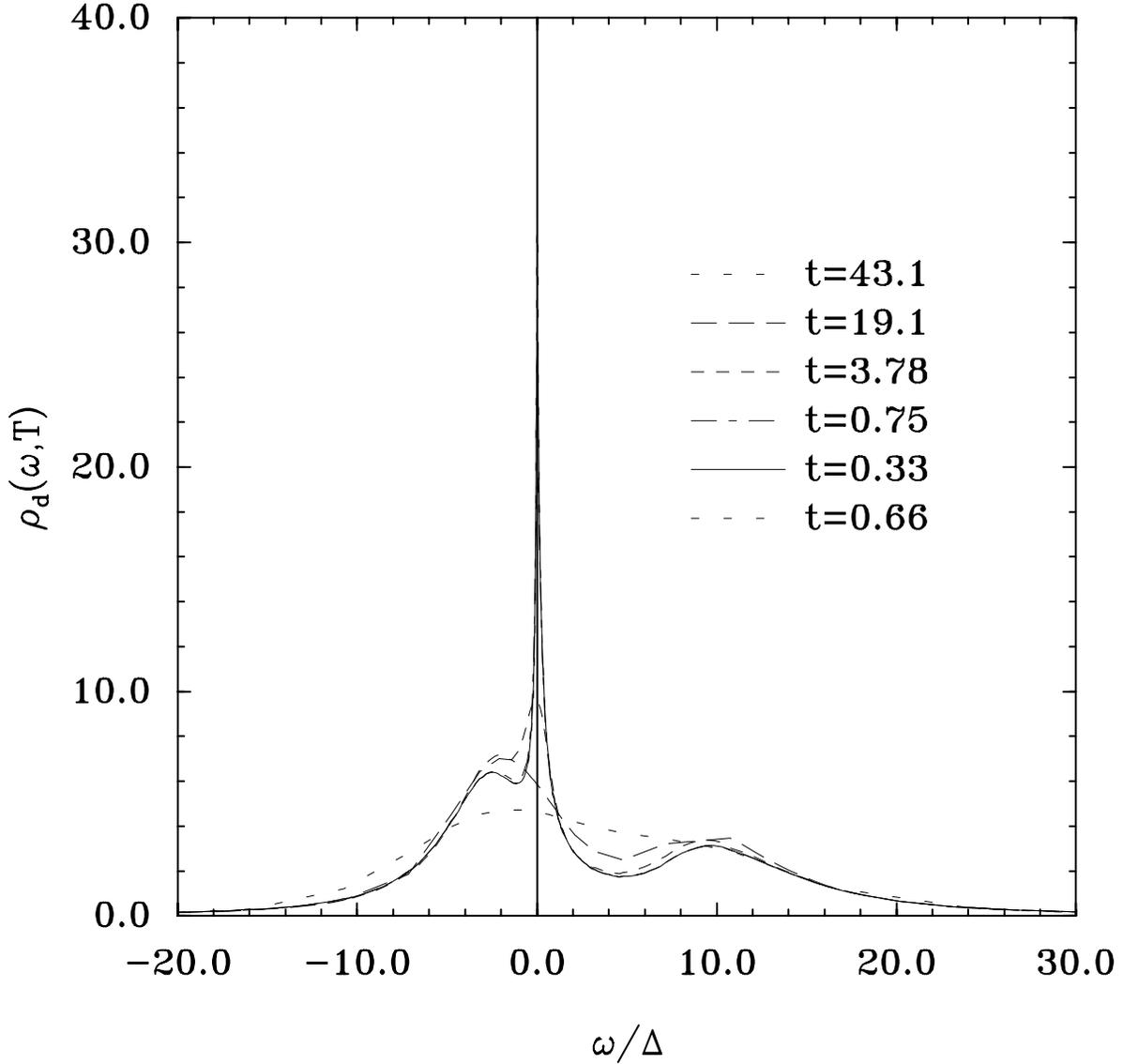}
\caption{The spectral density $\rho_{d}(\omega,T)$ in the Kondo regime
$\epsilon_{d}/\Delta = -3$ for several values of the reduced temperature
$t=T/T_{K}$ and for $U/\pi\Delta=4$.}
\label{figure2}
\end{figure}
\begin{figure}[t]
\includegraphics[width=\linewidth,clip]{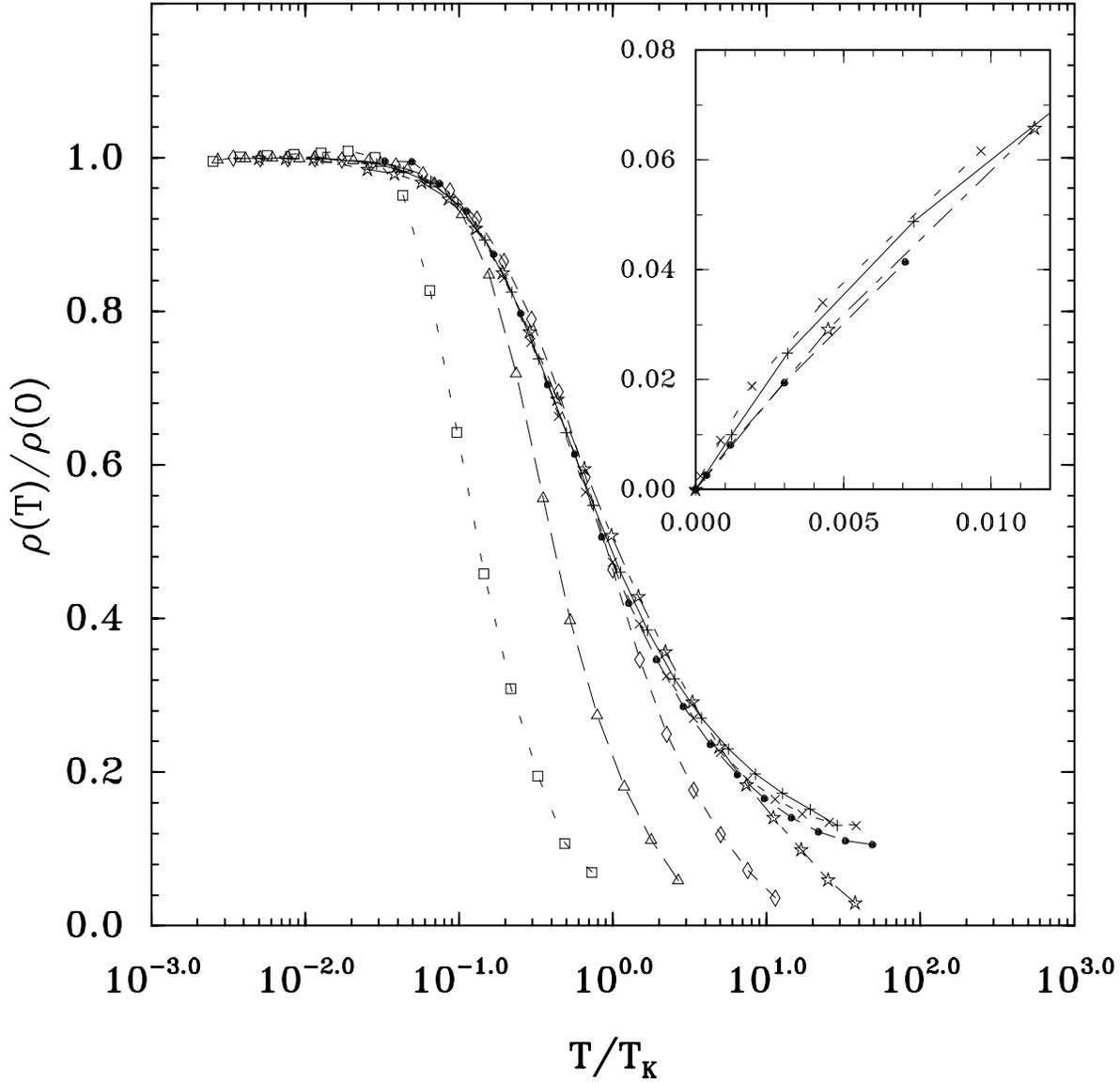}
\caption{The scaled resistivity $\rho(T)/\rho(0)$ over the
full temperature range in various regimes and for $U/\pi\Delta=4$, 
$\epsilon_{d}/\Delta=1$ ($\Box$), $\epsilon_{d}/\Delta=0$ 
($\triangle$), $\epsilon_{d}/\Delta=-1$ ($\diamond$),
$\epsilon_{d}/\Delta=-2$ ($\star$), $\epsilon_{d}/\Delta=-3$ ($+$),
$\epsilon_{d}/\Delta=-4$ ($\times$) and the symmetric case
($\bullet$). It should be noted that a  more appropriate energy scale
in the mixed valent and empty orbital regimes is $\Delta$.
The inset is for $(1-(\rho(T)/\rho(0)))$ versus $(T/T_{K})^{2}$ and
shows the $T^{2}$ Fermi liquid behaviour for $T << T_{K}$ in the Kondo regime.}
\label{figure3}
\end{figure}
\begin{figure}[t]
\includegraphics[width=\linewidth,clip]{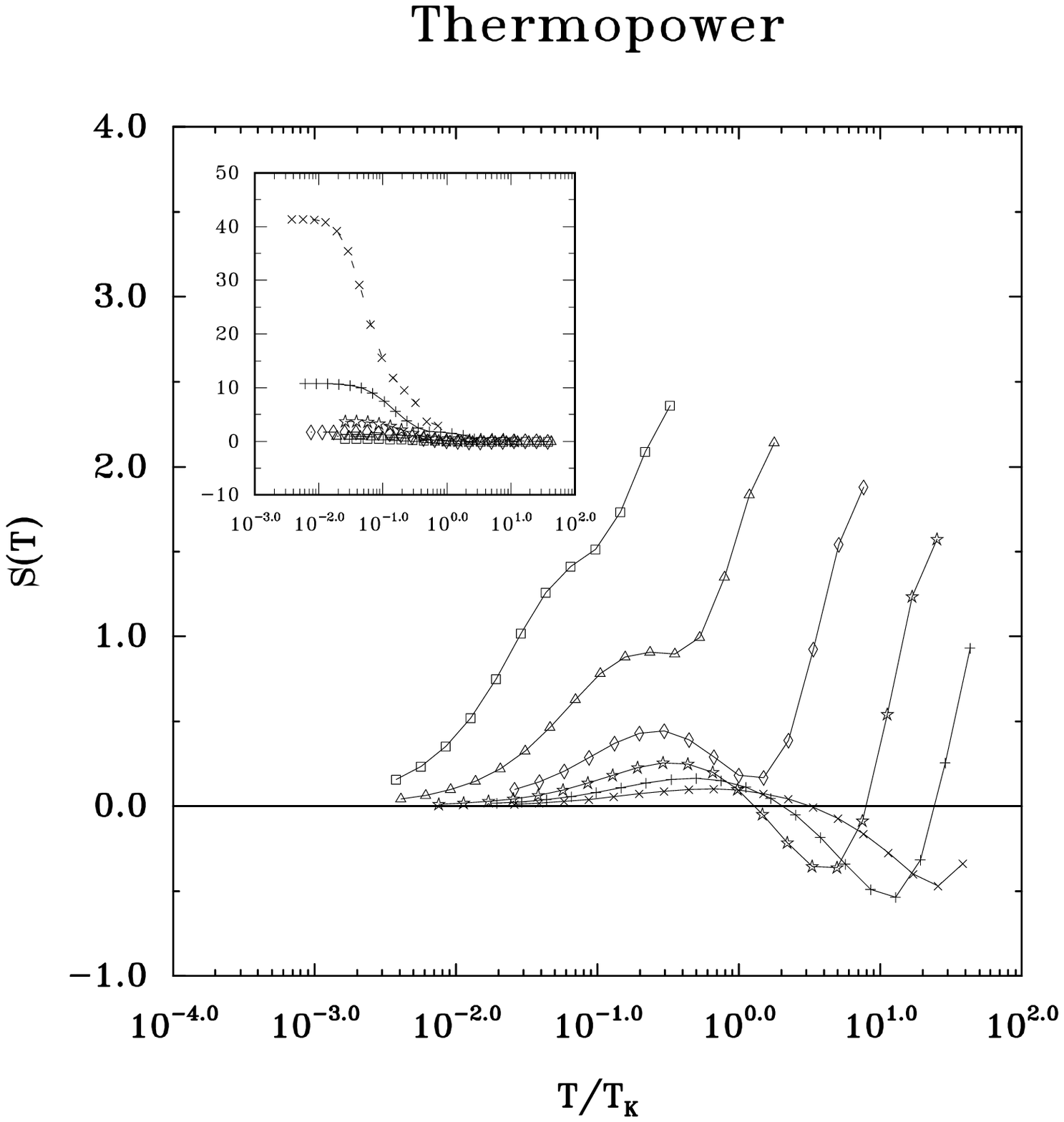}
\caption{The Thermopower $S(T)$ over the full temperature range
in various regimes and for $U/\pi\Delta=4$ 
$\epsilon_{d}/\Delta=+1$ ($\Box$), $\epsilon_{d}/\Delta=0$
($\triangle$), $\epsilon_{d}/\Delta=-1$ ($\diamond$),
$\epsilon_{d}/\Delta=-2$ ($\star$), $\epsilon_{d}/\Delta=-3$ ($+$),
$\epsilon_{d}/\Delta=-4$ ($\times$). The inset is for $S(T)T_{K}/T$
versus $T/T_{K}$ and shows the linear 
in T Fermi liquid behaviour for $T<<T_{K}$.}
\label{figure4}
\end{figure}
\begin{figure}[t]
\includegraphics[width=\linewidth,clip]{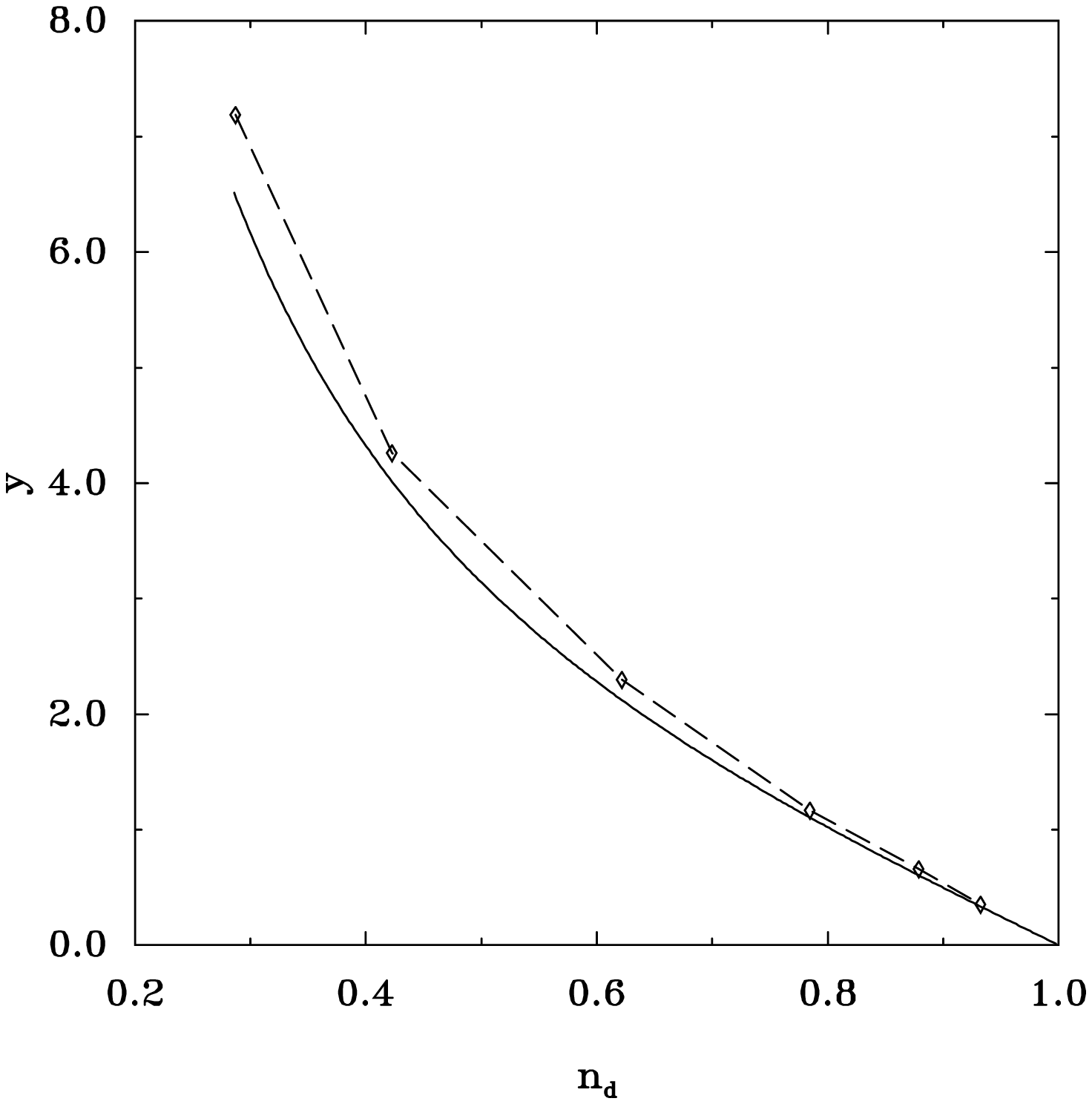}
\caption{{$\lim_{\thinspace T\rightarrow 0}\left\{ S(T) \over \gamma
T\right\}$ (dashed line with diamonds) and $\cot(\pi n_{d}/2)$ (solid
line) versus $n_{d}$ for the various parameter sets 
ranging from Kondo to mixed valent and empty orbital behaviour. The
two curves lie within $6\%$ of each other in all cases, except for the
empty orbital case $\epsilon_{d}/\Delta=+1$} where the error is 10\%.}
\label{figure5}
\end{figure}

A second Fermi liquid relation relates the linear coefficient of the
thermopower to the local level occupancy, $n_{d}(T=0)$, and the linear
coefficient of specific heat, $\gamma$, \cite{houghton87,kawakami87}
\begin{equation}
\lim_{\thinspace T\rightarrow 0}\left\{ {e S(T) \over {\gamma T}} \right\} 
= \pi\cot(\pi n_{d}(0)/2)
\end{equation}
This is a much more difficult Fermi liquid relation to satisfy,
because the local level occupancy $n_{d}(0)$ is obtained by
integrating the single--particle spectral density up to the Fermi
level, the linear coefficient of specific heat is obtained as a second
derivative $-\lim_{\thinspace T\rightarrow 0}\left\{ \partial^{2}F(T) \over
{\partial T^{2}}\right\}$ of the Free energy and $\lim_{\thinspace T\rightarrow
0}\left\{ S(T) \over T\right\}$
involves evaluating the transport integrals using the finite T
spectral densities $\rho_{d}(\omega,T)$. All these quantities are
evaluated using the numerical renormalization group method and the above
relation provides a very severe test of the accuracy of the method.

The thermopower is shown in Figure~\ref{figure4} for the mixed valent, empty orbital
and Kondo regimes.
The sign of the thermopower at low temperatures is determined
by the sign of $\left(\partial \rho(\omega,T) \over \partial
\omega\right)_{\omega=0}$ and is always positive for the range of
parameters considered. At higher temperatures the behaviour of $S(T)$
is more complicated. However, 
in the mixed valent and empty orbital regimes the single resonance
always lies above the Fermi level (see Figure~\ref{figure1}), and from equation (7) the
thermopower is positive at all temperatures. In the Kondo regime a
distinctive maximum appears in the thermopower at
$T\approx T_{K}$. At higher temperatures, the thermopower can become
negative and for sufficiently high temperatures positive again. This
bahaviour is associated with the temperature dependence of the
satellite peaks at $\epsilon_{d}$ and $\epsilon_{d}+U$ (see Figure~\ref{figure2}). 
The low temperature behaviour of the thermopower (inset to
Figure~\ref{figure4}) shows the expected linear in $T$ Fermi liquid behaviour and
one can use this to 
extract $\lim_{\thinspace T\rightarrow 0}\left\{ S(T) \over T\right\}$. By
calculating the linear coefficient of specific heat from the Free
energy and $n_d(T=0)$ from the spectral densities we can test the
Fermi liquid relation (10). This is shown in Figure~\ref{figure5}. 

It is seen that
$\lim_{\thinspace T\rightarrow 0}\left\{ S(T) \over \gamma T\right\}$
lies within $6\%$ of $\cot(\pi n_{d}/2)$ in nearly all cases. This agreement is
remarkable and shows clearly that the method is capable of giving close to exact
results for transport coefficients and finite temperature dynamic
properties in addition to the $T=0$ dynamics and the thermodynamics.
We have not made any of the sophisticated corrections
\cite{krishnamurthy80} to improve on the thermodynamics and only the
even energy shell spectral densities have been utilized, hence there is scope
for improvement. The characteristic peak at $T\approx
T_{K}$ in the thermopower in the 
Kondo regime is often referred to as the giant thermopower. This peak
can become very large, although to see this one needs to keep $n_d$
fixed and approach the Kondo regime by increasing U. In our calculations
$\epsilon_d$ was varied to approach the symmetric case for which the
thermopower is zero as can be seen by setting $n_{d}=1$ in equation
(10). From equation (10) it is clear that keeping $n_d$ fixed and 
increasing U leads to an exponential increase in the thermopower since
$\gamma \sim 1/T_{K} \sim e^{\alpha U}, \alpha > 0$ thus giving a giant
thermopower in the Kondo limit. The vanishing of the thermopower for
half--filling is an artifact of the simplified model used here which neglects
the non--resonant scattering of conduction electrons as discussed in \cite{zlatic74}.
An extension of the present calculations to include non--resonant scattering,
leads to enhanced thermopowers for half--filled systems and describes
the qualitative behaviour of the thermopower of concentrated Kondo compounds \cite{zlatic93}.

\section{\bf Conclusions} 
In this Letter we have presented new results for the thermopower and
resistivity of the Anderson model over the full temperature range of
interest and in several parameter regimes by using the numerical
renormalization group approach. The Fermi liquid relation for the
linear coefficient of the thermopower is satisfied with remarkable
accuracy, as is that for the $T^{2}$ coefficient of the resistivity.
The Hall coefficient, thermal conductivity and finite
temperature dynamic susceptibilities have also been calculated
and will be presented elsewhere \cite{costi93b,costi93c}. Our results
are in contrast to NCA 
calculations \cite{bickers87,qin91} which violate Fermi liquid
relations at low energies, and to perturbation in U calculations which
although accurate eventually break down for $U/\pi\Delta \ge 2.5$
\cite{horvatic87}. We are able to treat a much wider range of regimes
than the Monte Carlo approach \cite{jarrell91b} which has so far only been applied to
the symmetric case where the thermopower is zero and which often
builds in the sum rules which constitute independent tests within our
approach. The techniques presented here for extracting the spectral
densities of the Anderson model provide a new highly accurate
approach to the $d=\infty$ Hubbard model.

\subsection{Acknowledgments}
We acknowledge the support of an SERC grant, the Computational
Science Initiative for Computer equipment and Dr V. Zlati\'c for his
helpful comments.

\end{document}